\documentclass[aps,prl,twocolumn,epsf,epsfig,amsmath]{revtex4}

\usepackage{latexsym}
\usepackage{hyperref}
\usepackage{times}
\usepackage{graphicx}
\usepackage{amsmath}
\usepackage{multirow}

\usepackage{dcolumn}% Align table columns on decimal point
\usepackage{bm}% bold math
\usepackage{ textcomp }
\usepackage{color}
\usepackage{amssymb}
\usepackage{xcolor}
\usepackage{easyReview}
\usepackage{mathdots}
\usepackage{float}
\usepackage{lineno}
\usepackage{todonotes}
\usepackage{array}
\usepackage{subfigure}
\usepackage{layouts}
\usepackage{braket}
\usepackage[export]{adjustbox}
\usepackage{hyperref}
\newcommand{\beq}{\begin{eqnarray}}
\newcommand{\eeq}{\end{eqnarray}}

\newcommand{\bb}[1]{{{\mathbf #1}}}

\begin{document}

\title{Failure of Topological Invariants in Strongly Correlated Matter }
\author{Jinchao Zhao, Peizhi Mai, Barry  Bradlyn and Philip Phillips}
\affiliation{Department of Physics and Institute for Condensed Matter Theory,
University of Illinois
1110 W. Green Street, Urbana, IL 61801, U.S.A.}
\date{April 2023}

\begin{abstract}

We show exactly that standard `invariants' advocated to define topology for non-interacting systems deviate strongly from the Hall conductance whenever the excitation spectrum contains zeros of the single-particle Green function, $G$, as in general strongly correlated systems.  Namely, we show that if the chemical potential sits atop the valence band, the `invariant' changes without even accessing the conduction band but by simply traversing the band of zeros that might lie between the two bands.  Since such a process does not change the many-body ground state, the Hall conductance remains fixed.  This disconnect with the Hall conductance arises from the replacement of the Hamiltonian, $h(\bb k)$, with $G^{-1}$ in the current operator, thereby laying plain why perturbative arguments fail.
\end{abstract}

\maketitle

The stability of a gapped ground state against smooth deformations of the Hamiltonian that do not close a spectral gap is the cornerstone of topology. Such stability is captured by quantized invariants.  Key invariants that arise in topological systems are the Chern numbers.  While they appear as coefficients of the Chern-Simons Lagrangian, they have physical import as well.  For example, the first Chern number, $C_1$, is the coefficient,
\beq
\sigma_{\rm H}=C_1\frac{e^2}{h},
\label{eq:chern}
\eeq
of the Hall conductance~\cite{TKNN,Niu}%, $\sigma_H=\sigma_{\rm (xy)}$
.  As a topological invariant, $C_1$ can only change if 
%a gap closes, a case in point being the chemical potential moving from the top of the valence band to somewhere in the conduction band.  This would result in a unit increase in the Hall conductance. Consequently, 
the chemical potential crosses a band or more generally, if there are zero-energy excitations, measured with respect to the chemical potential. Any movement of the chemical potential within a spectral gap amounts to an adiabatic change of the system Hamiltonian, and so cannot change $C_1$.% Any movement of the chemical potential within the gap between the valence and conduction bands amounts to an infinitesimal change in the system Hamiltonian since no band crossings are encountered.\bbnote{I would alter the previous sentence to not mention bands at all, since this is more general. Something like ``Any movement of the chemical potential within a spectral gap amounts to an adiabatic change of the system Hamiltonian, and so cannot change $C_1$.} 
We will take such a change to be the paradigmatic definition of an infinitesimal deformation.

For computational purposes, it has become common to formulate Chern numbers in terms of single-particle Green functions.  Consider the commonly conceived invariant $N_3$\cite{volovik,gurarie} for the two-dimensional quantum anomalous Hall (QAH) insulator (also named as $N_2$ in Ref. \cite{wang2010topological,wang2012simplified})
\begin{equation}
\begin{split}
    N_3&=\frac{\epsilon_{\alpha\beta\gamma}}{6}\\
    &\mathrm{tr}\int_{-\infty}^\infty d\omega\int\frac{d^2\bb k}{(2\pi)^2}G^{-1}\partial_{k_\alpha}GG^{-1}\partial_{k_\beta}GG^{-1}\partial_{k_\gamma}G,
\end{split}
    \label{eq:n3}
\end{equation}
where $G(\omega,\bb k)$ is the zero temperature (single-particle) Green function in momentum space, $\alpha$, $\beta$ and $\gamma$ take values $0$, $1$ and $2$, such that $k_0=\omega$, and $k_1, k_2$ are components of the crystal momentum, and $\mathrm{tr}$ denotes the trace over the fermionic degrees of freedom of $G$.   For non-interacting electrons, $N_3$ reduces to the first Chern number $C_1$, or  equivalently the Thouless-Kohmoto-Nightingale-den Nijs (TKNN) \cite{TKNN} invariant.  That $N_3$ is invariant to small deformations of the Hamiltonian follows from substituting the infinitesimal,%\bbnote{suggestion: change the derivative subscript to $k_\alpha$ to be consistent with Eq. 2}
\begin{equation}
\begin{split}
    \delta(G\partial_{k_\alpha} G^{-1})&=\delta G\partial_{k_\alpha} G^{-1}- G\partial_{k_\alpha} (G^{-1}\delta G G^{-1})\\
    =&-G(\partial_{k_\alpha} G^{-1})\delta G G^{-1}-\partial_{k_\alpha} (\delta G) G^{-1},
\end{split}
\end{equation}

into the variation of $N_3$ which leads to a recasting of the resultant integrand as a total derivative. As the integral of a total derivative, $\delta N_3$ will naturally vanish for $\delta G$ continuously connected to zero (i.e.~for small deformations).  Consequently, $N_3$ is invariant to infinitesimal changes in the underlying Hamiltonian provided periodic boundary conditions are imposed. %\bbnote{There is a tension between this definition of invariance of $N_3$ under infinitesimal deformations, and the operational definition of infinitesimal deformation given in the first paragraph (since later we show changing the chemical potential can change $N_3$.} 

The utility of Eq.~(\ref{eq:n3}) is that only the Green function is required to evaluate $N_3$, rather than the full spectrum of the eigenstates as is typically needed to compute the Berry curvature or the TKNN invariant\cite{TKNN}.  Consequently, one may hope that Eq.~(\ref{eq:n3}) naturally applies to interacting systems.  However, when interactions are present, the Green function can vanish\cite{dzy} along a connected surface in momentum space for frequencies within the gap. This defines the Luttinger surface, which is a Mott fixed point under local perturbations\cite{HKnp2,rghk}.  What happens to $N_3$ when the chemical potential crosses such a surface?  If the ground state evolves continuously and the gap does not close, then the topological invariants of the ground state cannot change.  That is, $C_1$ should remain fixed.  However, it is known\cite{volovik,gurarie} that $N_3$ is sensitive to a zero or an edge-state (pole in the propagator) crossing the chemical potential. %, the general hope is that zeros are annihilated in lieu of poles thereby keeping the Chern number fixed.  
It is this sensitivity that underlies a recent claim that zeros are topological in the context of doped Mott insulators\cite{wagner}. In particular for models of fractional quantum Hall effect (FQHE), it has been shown that $N_3$ is in general not equal to the $C_1$\cite{gurarie2013topological}. Even more, pairs of fractional quantum Hall states with different Chern numbers (and hence different ground state topology) can be shown to have equal values of $N_3$. However, to our knowledge, the precise relationship between $N_3$ and $C_1$ as a function of chemical potential has not been established for an interacting system.

It is this loophole that we address in this paper.  For the Hatsugai-Kohmoto model\cite{HK,mai20221} with a topological non-trivial ground state, we use the exact Green function to show that even without closing the gap, $N_3$ changes when a band of zeros cross the chemical potential.  By definition, such a change constitutes an infinitesimal variation that does not close an energy gap, and hence there should be no change in topological invariants characterizing the ground state.  Consequently, we demonstrate explicitly that $N_3$ in Eq.~(\ref{eq:n3}) and $C_1$ are disconnected should zeros appear in the Green function. % We prove that the Hall conductance only reduces to $N_3$ provided there are no zeros of the Green function.  
In general for interacting systems, although $N_3$ is a topological property of the single-particle Green function, it does not necessarily encode a topological invariant of the ground state in contrast to previous claims\cite{volovik,gurarie,wagner}. %\bbnote{We should be a bit careful in the way this statement is formulated. Gurarie in his original paper seeemed to be aware of the fact that $N_3$ can change without closing a gap (i.e., in the absence of edge states), and tried to formulate a bulk-boundary correspondence between $N_3$ and the difference between the number of edge states and the number of edge zeros. Personally I don't think theres any reason to think edge zeros should be a robust indicator of anything, especially since they can be gotten rid of by, say, tuning the chemical potential! I made an attempt to clarify here and will add a short discussion to the end as well.} 
%The section{$N_3$ of an interacting topological system}

The computation of $N_3$ requires knowledge of the full single-particle Green function.  To this end, we adopt a model that affords an exact treatment of interaction and topology for the QAH effect\cite{mai2023topological}. For a square lattice with the orbitals positioned at lattice sites, the non-interacting part of a two-fold (spinful) Chern insulator can be written as,
\beq
H_0=\sum_{\bb k}c^\dagger_{\bb k}h(\bb k)c_{\bb k}=\sum_{\bb k}c^\dagger_{\bb k}
\begin{pmatrix}
h_{\text{QAH}}({\bb k}) & 0\\
0 & h_{\text{QAH}}({\bb k}) \label{QAH}
\end{pmatrix}c_{\bb k},
\eeq
where $c^\dagger = \{c^\dagger_{O_1,\uparrow} c^\dagger_{O_2,\uparrow},c^\dagger_{O_1,\downarrow} c^\dagger_{O_2,\downarrow} \}$ is a four-component spinor, and $O_{1/2}$ stands for different orbitals or sub-lattices, respectively. $h_{\text{QAH}}({\bb k})=h_\alpha({\bb k})\tau^\alpha$ describes a $2\times2$ QAH Hamiltonian for each spin. This Hamiltonian can be diagonalized under a unitary transformation into $h(\bb k)=V(\bb k)\mathrm{diag}(\varepsilon_{-,{\bb k}},\varepsilon_{-,{\bb k}},\varepsilon_{+,{\bb k}},\varepsilon_{+,{\bb k}})V^\dagger(\bb k)$ 
%\bbnote{$V(\bb k)$ should be given explicitly in the appendix}
where upper ($+$) and lower ($-$) bands are given by
\beq
\varepsilon_{\pm,{\bb k}}=h_{0}({\bb k})\pm\sqrt{h_{x}^2({\bb k})+h_{y}^2({\bb k})+h_{z}^2({\bb k})}.
\eeq
%For $2>M>0$ (or $-2<M<0$), the half-filled system is a QSH insulator\cite{bhz} with Chern number $C_0=C_{\uparrow0}+C_{\downarrow0}=0$ and spin Chern number $C_{s0}=C_{\uparrow0}-C_{\downarrow0}= 2$ (or $-2$) related to the spin Hall conductance\cite{bhz,Qi1}, where the subscript $0$ means no interaction. $\big|M\big|>2$ corresponds to a topologically trivial band insulator with Chern numbers $C_{s0}=C_0=0$. 
Electrons with opposite spin have the same dispersion and chirality. This momentum space basis is not destroyed under the local-in-momentum Hatsugai-Kohmoto (HK) interaction that includes Mottness\cite{HK,mai2023topological,HKnp1,HKnp2,hksc} 
\beq
\begin{aligned}
H&_{\rm{QAH-HK}}=\sum_{{\bb k},\sigma}\big[(\varepsilon_{+,{\bb k}}-\mu)n_{+,{\bb k},\sigma}+(\varepsilon_{-,{\bb k}}-\mu)n_{-,{\bb k},\sigma}\big]
\\&+U\sum_{\bb k}(n_{+,{\bb k},\uparrow}+n_{-,{\bb k},\uparrow})(n_{+,{\bb k},\downarrow}+n_{-,{\bb k},\downarrow}). \label{HHK}
\end{aligned}
\eeq
The interaction term is rotational symmetric under the unitary transform $V(\bb k)$ since $n_{+,{\bb k},\sigma}+n_{-,{\bb k},\sigma}$ is a trace in either the orbital or band basis. The exact Green function in the band basis,
\begin{equation}
\begin{aligned}
G_{\pm,{\bb k},\sigma}(\omega)=&\frac{\langle(1- n_{+,{\bb k},\bar{\sigma}})(1-n_{-,{\bb k},\bar{\sigma}})\rangle}{\omega+\mu-\varepsilon_{\pm,{\bb k}}} \\
&+ \frac{\langle n_{+,{\bb k},\bar{\sigma}}+n_{-,{\bb k},\bar{\sigma}}-2n_{+,{\bb k},\bar{\sigma}}n_{-,{\bb k},\bar{\sigma}}\rangle}{\omega+\mu-(\varepsilon_{\pm,{\bb k}}+U)}\\
&+ \frac{\langle n_{+,{\bb k},\bar{\sigma}}n_{-,{\bb k},\bar{\sigma}}\rangle}{\omega+\mu-(\varepsilon_{\pm,{\bb k}}+2U)},\\
\end{aligned}
\label{eq:green}
\end{equation}
has 6 poles at any given momentum. However, only some of them have a non-vanishing weight in the insulating state at $U\gg W$ (bandwidth) sufficiently large $U$%\bbnote{here you say sufficiently large, but later you say $U\gg W$. Is it the same constraint in both places? If so should say it in the same way}.
At quarter-filling, the degenerate $\varepsilon_-$ band is singly occupied, thus $\langle n_{-,{\bf k},\uparrow}\rangle=\langle n_{-,{\bf k},\downarrow}\rangle=1/2$. The $\varepsilon_+$ band remains empty for both spin, $\langle n_{+,{\bf k},\sigma}\rangle=0$ and $\langle n_{+,{\bf k},\bar{\sigma}}n_{-,{\bf k},\bar{\sigma}}\rangle=0$. Thus, the poles at $\varepsilon_{\pm,{\bb k}}-\mu+2U$ have zero weight. 

At half-filling and $U\gg W$, the ground state always occupies both $\varepsilon_\pm$ with the same spin, $\langle n_{-,{\bf k},\sigma}\rangle=\langle n_{+,{\bf k},\sigma}\rangle=1/2$ and $\langle n_{+,{\bf k},\sigma}n_{-,{\bf k},\sigma}\rangle=\frac{1}{2}$. Thus the poles at $\varepsilon_{\pm,{\bf k}}-\mu+U$ have zero weight. The remaining 4 poles all have the same weight of $1/2$, generating the zero branches located at 
%has 4 poles at a given momentum, as well as 2 zeroes located at 
the poles of the self-energy,
\begin{equation}
\Sigma_{\pm, \bb k,\sigma}(\omega)=U+\frac{U^2}{\omega+\mu-\varepsilon_{\pm ,\bb k}-U}.
\end{equation}
The position of the 4 poles relative to the chemical potential defines the electron filling. In the case of half-filling, the lower two poles located at $\varepsilon_{\pm,{\bb k}}-\mu$ lie below the chemical potential, while $\varepsilon_{\pm,{\bb k}}-\mu+2U$ lies above, thereby maintaining the gapped Mott state.

%Although the HK interaction correlates spin-up and spin-down, its Green function does not and remains block diagonal in terms of the spin degrees of freedom. Therefore, the topological invariant $N_{3\sigma}$ for each spin is well defined by the $2\times2$ Green function in each spin block.
%In the rest of this paper, $N_3$ refers to $N_3$ and $N_{3\downarrow}=-N_3$ \bbnote{I would prefer avoiding this redefinition and using $N_3$ throughout the rest of the paper. As is, its very confusing for someone that approaches papers in the traditional way of first reading the abstract and intro, then the conclusion, and then the body of the text.}. 
According to a previous analysis\cite{mai2023topological} on the topology of this model, we know that both the QAH-HK and QAH-Hubbard models predict a topologically trivial phase at half-filling when the interactions dominate. There is a topological phase transition from half-filling to quarter-filling, leading to a topological Mott insulator at quarter-filling with $C_1=1$, which is half the Chern number of the band insulator at half filling
.

\begin{figure}[htbp]
    \centering
    \subfigure[$N_3=0$]{\includegraphics[width=(\textwidth-\columnsep)/7]{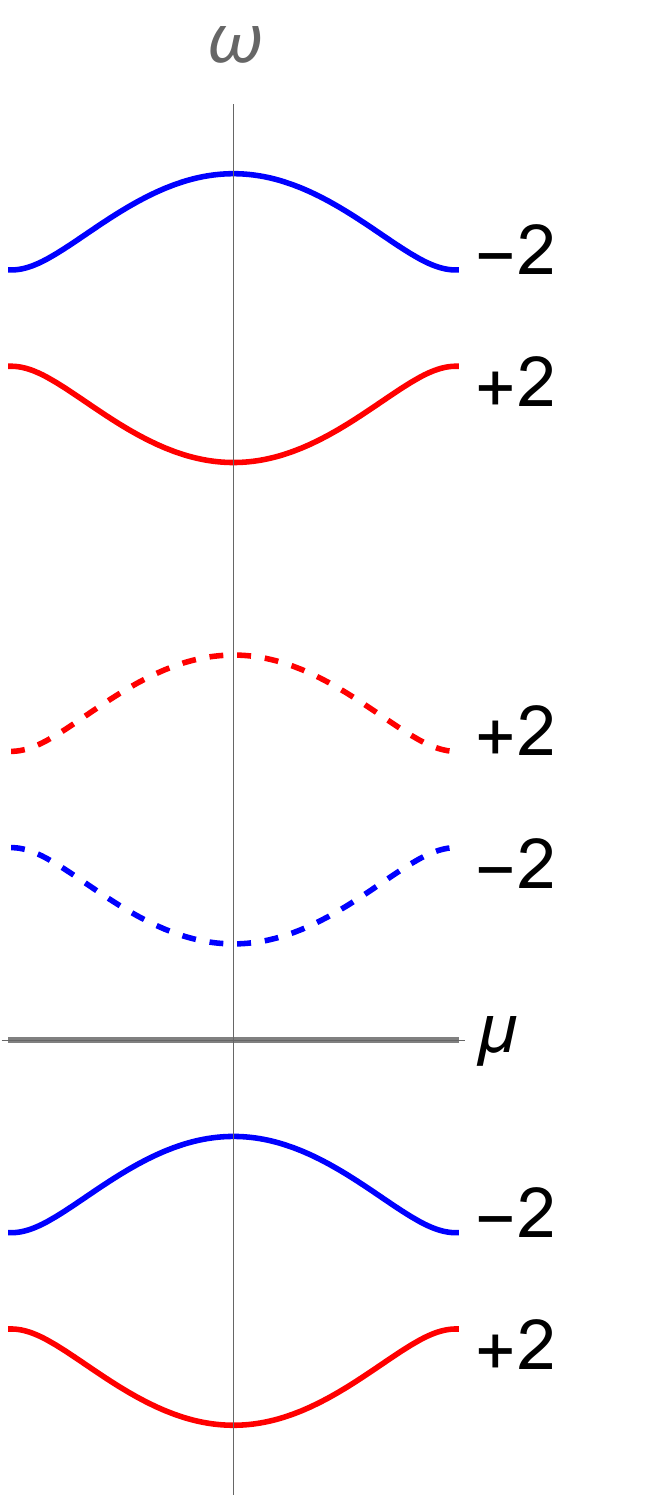}}
    \quad
    \subfigure[$N_3=-2$]{\includegraphics[width=(\textwidth-\columnsep)/7]{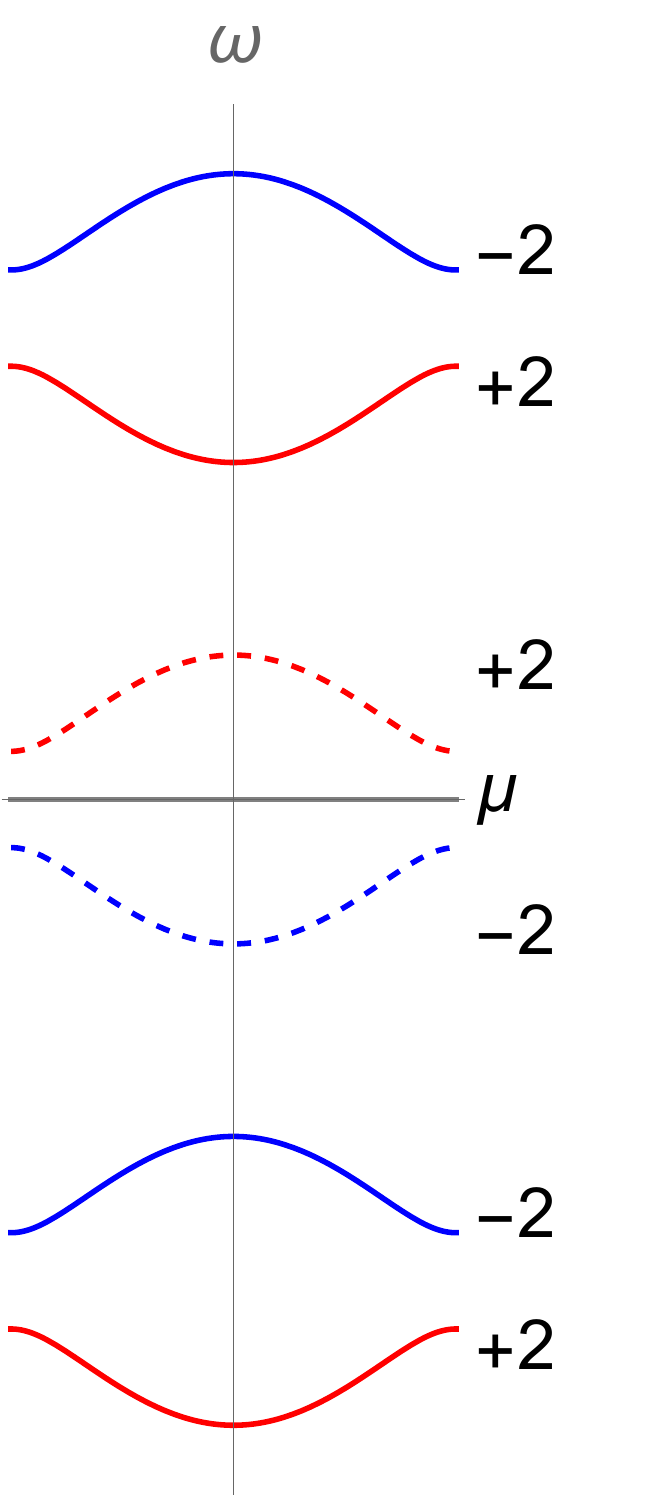}}
    \quad
    \subfigure[$N_3$ undefined]{\includegraphics[width=(\textwidth-\columnsep)/7]{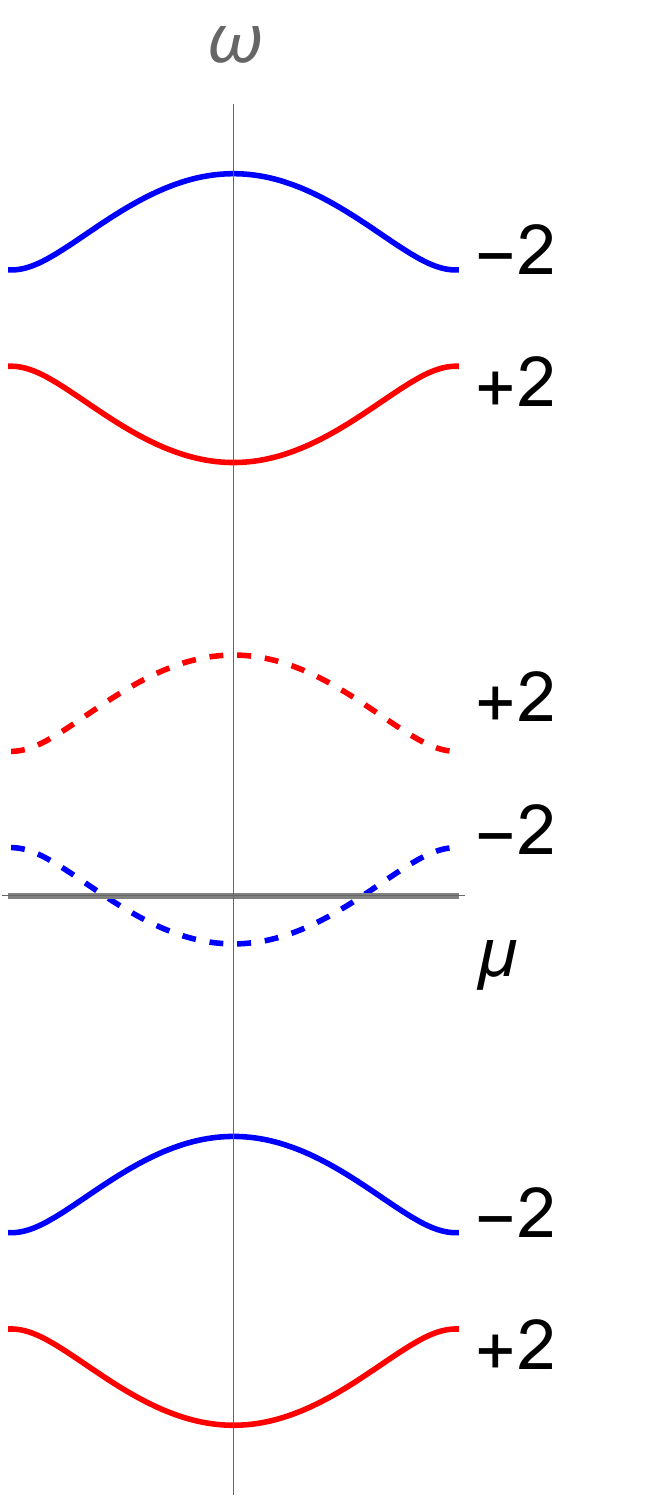}}
    \quad
    \caption{The pole structure for the Green function from Eq.~(\ref{eq:green}) at half-filling with $U\gg W$ (bandwidth). The solid lines represent the poles, the dashed line represents the zeroes. The numbers next to the curves are the corresponding contribution to $N_3$ of that particular band. Note that all three of these configurations represent the same gapped half-filling ground state, while the $N_3$ for each setup is $N_3=0,-2,$ or undefined. Here we use the Haldane model\cite{haldane} as an example for Eq.~(\ref{HHK}) with an HK interaction to construct the band dispersion. 
    }
    \label{fig:QSHband}
\end{figure}

At any filling with a gap, such as half-filling where $U$ sets the gap scale, we can shift the chemical potential $\mu$ inside this gap without affecting the many-body ground state. 
%\bbnote{A plot of the excitation energy (i.e. spectral gap, rather than single-particle gap) would help here to illustrate the point}.  
As this constitutes an infinitesimal variation of the Hamiltonian, there should be no change in the topology. However, this shift of $\mu$ drastically changes the value of $N_3$ due to the location of zeroes, as shown in Fig.~\ref{fig:QSHband}. At half-filling, $\braket{n_{\pm,\bb k,\bar\sigma}}=1/2$ for both of the spin as well as the upper($+$) and lower($-$) topological bands. All the spinful bands of zeroes or poles below the chemical potential contribute a $\pm2$ to $N_3$ as labeled in Fig.~(\ref{fig:QSHband}). The zero bands locate at $\varepsilon_{\pm ,\bb k}-\mu+U$. If the branches of the zeros are located on the same side of the chemical potential (Fig.~\ref{fig:QSHband}(a)), $N_3=0$. In the vicinity of the symmetry point, $\mu=U$, the chemical potential is located between the two zero branches (Fig.~\ref{fig:QSHband}(b)), giving rise to a non-zero $N_3=-2$. When the chemical potential passes through the zeroes band (Fig.~\ref{fig:QSHband}(c)), $N_3$ diverges as if the system is in a metallic state. 

This seems to give rise to a contradiction if we expect $N_3$ to be proportional to the Chern number (equivalently, the Hall conductance).
%as any change in $N_3$ must be correlated with a change in the Hall conductance or equivalently the Hall number. 
That is, there seems to be a change in the topological invariant without changing the many-body ground state.  A similar change in the Luttinger count has been noted previously\cite{rosch,dave}, because moving the chemical potential in the gap changes the positions of the zeros but ultimately cannot change the filling.  It is for this reason that it has been correctly argued that the Luttinger count, which counts zeros and poles, does not enumerate the charge density in generic interacting systems. Similarly, we have shown here explicitly that $N_3$ is counting both zeros and poles of the Green function and hence does not enumerate the Chern number in general. This derivation %\bbnote{Do you mean ``This derivation''?} 
could also apply to the quantum spin Hall (QSH) system\cite{bhz,kanemele1,kanemele2} with strong interactions where a similar interaction-induced topological phase is observed\cite{mai20221}. 

\begin{figure}[htbp]
    \centering
    \subfigure{\includegraphics[width=(\textwidth-\columnsep)/3]{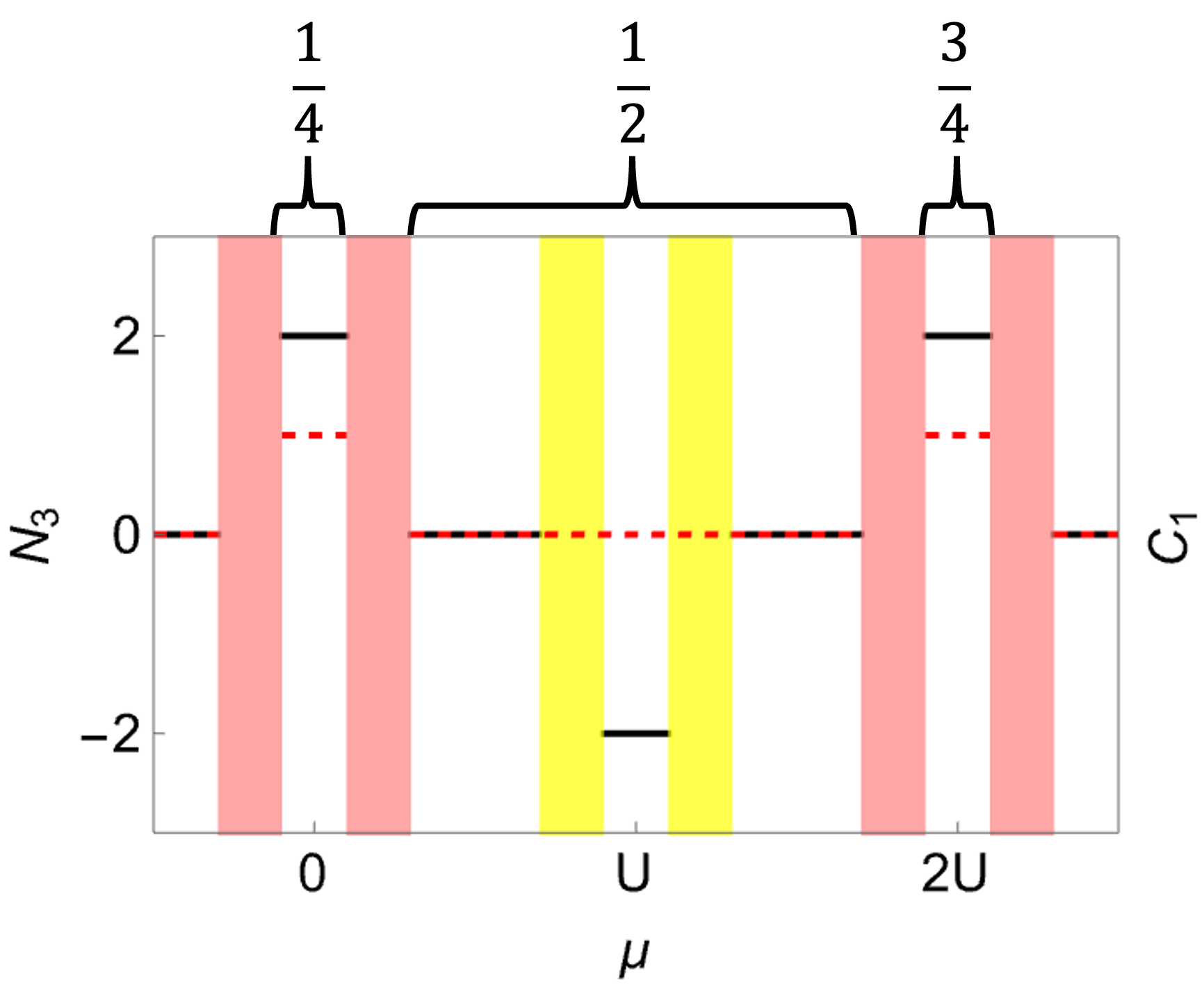}}
    \caption{The change of $N_3$ (Solid Black line) according to Eq.~(\ref{eq:n3}) and $C_1$ (Dashed Red line) according to Eq.~(\ref{eq:response}) for a QAH-HK model with $U\gg W$ as a function of the chemical potential $\mu$. The fillings are labeled at the top. Inside the red regions, both $N_3$ or $C_1$ are undefined due to the crossing of poles (metallic state); inside the yellow regions, $N_3$ is undefined due to the crossing of zeros.}
    \label{fig:phasediag}
\end{figure}
%\section{the Hall conductance}
To address this conundrum, we compute the Hall conductance directly and establish when it is permissible for it to be recast as $N_3$. The advantage of the HK model is that the interactions preserve the center of mass and $U$ does not have any dependence on momentum. Thus, the current operator in the orbital basis %\bbnote{the orbital indices here need to be fixed}
\begin{equation}
    \bb J(\bb q)=\frac{1}{\sqrt{V}}\sum_{\bb k}c^\dagger_{\bb k-\bb q/2}\frac{\partial h(\bb k)}{\partial \bb k}c_{\bb k+\bb q/2},
    \label{eq:current}
\end{equation}
can be taken to be unchanged from its non-interacting form (See Appendix), where $h(\bb k)$ is the $4\times4$ non-interacting Hamiltonian defined in Eq.~(\ref{QAH}).%\bbnote{This is not entirely obvious to me, since the HK interaction is diagonal in the band basis whereas the current operator is not. I.e., $\partial_k h$ need not commute with $V(\bb k)$. It is worth showing this explicitly. Its also worth noting that this form of the $q$-dependent current operator is only necessarily correct to order $\mathcal{O}(q)$ if $h(\bb k)$ is not quadratic in momentum (a student and I are writing a separate paper about this more generally). This is a bit tangential here though, since I think at the end of the day we only need the $q\rightarxrow 0$ current.}
We substitute this current operator into the Kubo formula\cite{kubo1957statistical} and obtain the current-current response function at finite temperature
\begin{equation}\label{eq:R_before_wick}
\begin{split}
    R_{\alpha\beta}&(q,\tau)=\braket{T[J_\alpha(q,\tau)J_\beta(-q,0)]}\\
    &\quad\quad=\frac{1}{V}\sum_{k,k'}\frac{\partial h^{ab}(\bb k)}{\partial k_\alpha}\frac{\partial h^{cd}(\bb k')}{\partial k'_\beta}\\
    &\braket{T[c^\dagger_{k-q/2,a}(\tau)c_{k+q/2,b}(\tau)c^\dagger_{k'+q/2,c}c_{k'-q/2,d}]},\\
\end{split}    
\end{equation}
where $\alpha$ and $\beta$ represent real-space directions and $a,b,c,d$ are orbital and spin indices.
Since the HK interaction does not mix momentum, the 4-fermion correlation function can be calculated according to Wick's theorem\cite{rghk}.  We find that
%\cite{HKRG}
\begin{equation}
\begin{split}
    &\braket{T[c^\dagger_{k-q/2,a}(\tau)c_{k+q/2,b}(\tau)c^\dagger_{k'+q/2,c}c_{k'-q/2,d}]}\\
    &=\braket{c^\dagger_{k-q/2,a}(\tau)c_{k'-q/2,d}}\braket{c_{k+q/2,b}(\tau)c^\dagger_{k'+q/2,c}}.
\end{split}
\end{equation}
The Fourier transform of the current-current response function gives $j_\alpha(q,\omega)=R_{\alpha\beta}(q,\omega)A_\beta(q,\omega)$. The conductivity is thus given via analytical continuation $\sigma_{\alpha\beta}(\omega)=\lim_{q\rightarrow 0}\frac{1}{i\omega} R_{\alpha\beta}(q,i\nu_r\rightarrow\omega+i\eta)$ with
%\begin{widetext}
\begin{equation}
\begin{split}
     &R_{\alpha\beta}(q,i\nu_r)=\frac{k_BT}{V}\sum_{k,n}\\
     &\mathrm{Tr}\left[\frac{\partial h(\bb k)}{\partial k_\alpha}G(k+q/2,\omega_n)\frac{\partial h(\bb k)}{\partial k_\beta}G(k-q/2,\omega_n-\nu_r)\right].
     \label{eq:response}
\end{split}
\end{equation}
%\end{widetext}
For a non-interacting system, $h(\bb k)$ in Eq.~(\ref{eq:R_before_wick}) can be replaced by $G^{-1}$ which will bring the Hall conductance into the form of $N_3$.  However, for an interacting system, no such correspondence can be made; in general  %Clearly the condition for the replacement of {$h(\bb k)$ with $G^{-1}$} is that $\frac{\partial\Sigma(k,\omega)}{\partial k}$ does not generate any contribution. The deformation of the self-energy into a trivial flat band could be applied only if there are no poles in the full Brillouin zone. 
\begin{equation}\label{eq:GFkderiv}
    \frac{\partial G^{-1}(\bb k)}{\partial k_\alpha}=\frac{\partial h(\bb k)}{\partial k_\alpha}+\frac{\partial \Sigma(\bb k)}{\partial k_\alpha},
\end{equation}
 because the presence of the self-energy in the Green function introduces added momentum dependence. A non-trivial $\Sigma(\bb k)$ with a band of poles (yielding the band of zeros in the Green function as shown in Fig.~\ref{fig:QSHband}(b)) gives rise to the non-zero contribution to $N_3$. Also, $\Sigma(\bb k)$ diverges at the Luttinger surface, accounting for the undefined $N_3$ in Fig.~\ref{fig:QSHband}(c). Hence, for any interacting model with a pole in its self-energy, replacing $h(\bb k)$ with $G^{-1}$ fails. 
%Unfortunately, this condition is only true for local real-space models in which there is no current.  Hence, for any interacting model with kinetic energy, replacing {$h(\bb k)$ with $G^{-1}$} fails. 
As a consequence, there will be a general disconnect between $N_3$ with the Hall conductance whenever zeros exist.  A recent derivation of the Hall conductance  using diagrammatic perturbation theory  (which inherently assumes adiabatic continuity with the non-interacting limit) purports to derive an equivalence between $N_3$ and $C_1$~\cite{blason}.   As zeros in the Green function indicate that the self-energy diverges, no such adiabatic continuity exists and hence any correspondence between $N_3$ and $C_1$ fails based on perturbative arguments.  This is consistent with two prior results.  First, the breakdown of Luttinger's theorem for interacting systems stems has been tied to the non-existence of the Luttinger-Ward functional on account of poles in the self energy\cite{dave}.  Second, the disconnect between $N_3$ and $C_1$ for fractional quantum Hall states---which are not perturbatively connected to non-interacting topological phases---was pointed out in Ref.~\cite{gurarie2013topological}.  Consequently, anytime there is a breakdown of perturbation theory, $N_3$ and $C_1$ cannot be diretly related.
%In a recent derivation of the Hall conductance\cite{blason} in terms of the Green function, it was pointed out that the Luttinger-Ward identity requires that  % We note that within the approximations of dynamical mean field theory (DMFT), the momentum dependence of the self-energy is usually ignored. In particular, the DMFT approximation consists of self-consistently including the dynamical (frequency-dependent), spatially uniform ($\mathbf{k}$-independent) renormalization of the self energy~\cite{georges1996dynamical}. Within DMFT one would thus find that the second term on the right-hand side of Eq.~\eqref{eq:GFkderiv} vanishes, yielding $N_3=C_1$.
%\bbnote{Just wanted to check that we indeed wanted to comment out the two sentences about DMFT}

In all such cases, the Hall conductance will have to be computed directly from the Kubo formula, or equivalently by integrating the Berry curvature as a function of twisted boundary conditions~\cite{Niu}.  We illustrate this here with a computation of the Hall conductance directly from Eq.(\ref{eq:green}).  The full details are provided in the Appendix.
We define the Chern number of the ground state according to Eq.~(\ref{eq:chern}), where the ground state is taken to be the zero-temperature limit of a thermal state to account for the spin degeneracy. At quarter-filling, the Hall conductance $C_1=1$  is halved compared with the non-interacting two-fold QAH result $C_1=2$. At half-filling, the Hall conductance remains zero as long as no pole of the Green function crosses the chemical potential. ther%\bbnote{What process?}.
To illustrate the deviation of $N_3$ from the Hall conductance $C_1$, we plot their values as a function of the chemical potential $\mu$ in Fig.~\ref{fig:phasediag}. Besides the conflict between a non-zero $N_3$ and a vanishing $C_1$ at half-filling, we observe an additional difference by a factor of $2$ at quarter-filling between them. This difference of factor is similar to the deviation between $N_3$ and $C_1$ in FQHE\cite{gurarie2013topological,Niu} caused by the ground state degeneracy. %\bbnote{Im not sure what this comment about the ground state degeneracy means. Gurarie related $N_3$ in the FQH to the weight of the electron operator, which in general is disconnected from the ground state degeneracy}.
Thus, neither the trivial phase at half-filling nor the topological phase at quarter-filling could be captured accurately by $N_3$. This invariant fails to capture properties of the ground state that are robust to perturbations of the Hamiltonian, both qualitatively and quantitatively.
%\section{final remarks}
We have thus shown that the deviation of $N_3$ from $C_1$ stems from poles in the self-energy or equivalently zeros of the single-particle Green function.  A similar problem occurs for the Luttinger count,
\beq
n=2\int_{\rm Re G(p,\omega=0)>0}\frac{d^d p}{(2\pi)^d},
\eeq
which makes no distinction between the mechanisms for $\rm Re G(p,\omega)$ crossing the real axis.  There is now ample evidence\cite{rosch,dave,bedell} that it is zeros that disconnect the Luttinger count from the physical particle density.  At play here is a similar trend: any movement of the chemical potential within the gap changes the Luttinger count but ultimately should not change the physical charge density.  This is not surprising as the Luttinger count is reducible to the analogous expression for $N_3$ with just a single product $\partial G^{-1} G$, thereby defining $N_1$\cite{gurarie}. It was shown in Ref.~\cite{manmana2012topological} that two $1+1$-dimensional interacting systems with unequal $N_1$ could nevertheless possess topologically equivalent ground states.  Taken together, we see that all generalized invariants of the form, $N_\ell$ are disconnected from the physics of the many-body ground state because of the zeros of the single-particle Green function. For both $N_1$ and $N_3$, this discrepancy arises precisely when the \emph{single-particle} Green function fails to accurately capture properties of the \emph{many-body} ground state; the emergence of Green function zeros signifies the importance of multi-particle spectral weight. The charge density and Hall conductance, being properties of the ground state and \emph{not} properties of single-particle excitation, encode physics beyond the single-particle Green function. Finally, we note that Refs.~\cite{gurarie,wagner} showed that at the interface between two systems across which $N_3$ jumps by $\Delta N_3$ with no other differing topological invariant, there will be $\Delta N_3$ zeros in the boundary Green function. While this result is certainly correct and encodes topological properties of the single-particle Green function, our work here calls into question the significance of this result for ground-state topological properties. In particular, we have shown here that $N_3$ can jump at an interface where the chemical potential changes smoothly while remaining in the bulk gap. Although the single-particle Green function will develop boundary zeros, we have shown that robust observables computed from the many-body ground state cannot change across the interface. In order to reconcile these observations, what is needed is an analysis of higher-order correlation functions to reinstate the connection between ground state topology and robust observable~\cite{soldini2022interacting}.

\textbf{Acknowledgements} 
We thank J. Cano and N. Wagner for clarifying exchanges.  PWP and JZ acknowledge NSF DMR-2111379 for partial funding for work on the HK model. This work was also supported by the Center for Quantum Sensing and Quantum Materials, a DOE Energy Frontier Research Center, grant DE-SC0021238 (P.M., B.B., and P.W.P.). B.B. received additional support from NSF DMR-1945058 for his general work on topology.

%For the QSH-HK model we introduced at the beginning, its Green function $G(k,\omega)$ and its non-interacting Hamiltonian $h(\bb k)$ do not have a spin-mixing component. The Chern number for each spin, $C_\uparrow$, and $C_\downarrow$, is thus well defined. 

\bibliography{chern}
\section{Appendix}
\subsection{Current operator of HK model}
Here we provide a detailed derivation of the current operator of the HK model. We start with a generalized form of the HK Hamiltonian in a band basis,
\begin{equation}
\begin{split}
    H=&H_0+H_I\\
    =&\sum_{\bb k,\sigma}c^\dagger_{\bb k,a\sigma}h_{\sigma}^{ab}(\bb k)c_{\bb k,b\sigma}\\
    +&\sum_{\bb k}A^{abcd}(\bb k)c^\dagger_{\bb k,a\uparrow}c_{\bb k,b\uparrow}c^\dagger_{\bb k,c\downarrow}c_{\bb k,d\downarrow},
\end{split}
\end{equation}
where $h_{\sigma}(\bb k)$ is the $2\times2$ QAH Hamiltonian for each spin, $a,b,c,d$ are orbital indices which take value from $O_1$ or $O_2$.
The current satisfies the continuity equation,
\begin{equation}
    \frac{\partial \rho(x)}{\partial t}+\bb\nabla\cdot \bb J(x)=0.
\end{equation}
The density operator for origin-localized orbitals can be Fourier transformed into
%\bbnote{Pedantic point, but this is actually $\rho(-q)$ (I had it wrong in my handwritten notes). It wont change the final result of course.}
\begin{equation}
    \rho(q)=\frac{1}{\sqrt{V}}\sum_{\bb k,a,\sigma}c^\dagger_{\bb k,a\sigma}c_{\bb k+\bb q,a\sigma}.
\end{equation}
The continuity equation then yields
\begin{equation}
    \bb q\cdot\bb J(q)=[\rho(q),H]=[\rho(q),H_0]+[\rho(q),H_I],
\end{equation}
where the first term is the non-interacting current operator Eq.~(\ref{eq:current}). We focus on the contribution from the second term. According to the general Fermion commutation relation
\begin{equation}
\begin{split}
    &\sum_{\bb k',e}[c^\dagger_{\bb k',e\sigma}c_{\bb k'+\bb q,e\sigma},c^\dagger_{\bb k,a\sigma}c_{\bb k,b\sigma}]\\
    &=c^\dagger_{\bb k-\bb q,a\sigma}c_{\bb k,b\sigma}-c^\dagger_{\bb k,a\sigma}c_{\bb k+\bb q,b\sigma}\\
    &=-\bb q\cdot\nabla_{\bb k}c^\dagger_{\bb k,a\sigma}c_{\bb k,b\sigma}-c^\dagger_{\bb k,a\sigma}\bb q\cdot\nabla_{\bb k}c_{\bb k,b\sigma}\\
    &=-\bb q\cdot\nabla_{\bb k}(c^\dagger_{\bb k,a\sigma}c_{\bb k,b\sigma}),
\end{split}
\end{equation}
where we have expanded the operator to linear order in $q$ and neglected higher order terms $O(q^2)$ since we will take the $q\rightarrow0$ limit in the Kubo formula.
The commutator between the density operator and the interaction is
\begin{equation}
\begin{split}
    &\sum_{\bb k}A^{abcd}(\bb k)\sum_{\bb k',e,\sigma}[c^\dagger_{\bb k',e\sigma}c_{\bb k'+\bb q,e\sigma},c^\dagger_{\bb k,a\uparrow}c_{\bb k,b\uparrow}c^\dagger_{\bb k,c\downarrow}c_{\bb k,d\downarrow}]\\
    %=&\sum_{\bb k}A^{abcd}(\bb k)\sum_{\bb k',e}[c^\dagger_{\bb k,a\uparrow}c_{\bb k,b\uparrow},c^\dagger_{\bb k'+q,e\uparrow}c_{\bb k',e\uparrow}]c^\dagger_{\bb k,c\downarrow}c_{\bb k,d\downarrow}\\
    %&\qquad+c^\dagger_{\bb k,a\uparrow}c_{\bb k,b\uparrow}[c^\dagger_{\bb k,c\downarrow}c_{\bb k,d\downarrow},c^\dagger_{\bb k'+q,e\downarrow}c_{\bb k',e\downarrow}]\\
    =&-\sum_{\bb k}A^{abcd}(\bb k)\left(\bb q\cdot\nabla_{\bb k}(c^\dagger_{\bb k,a\uparrow}c_{\bb k,b\uparrow})c^\dagger_{\bb k,c\downarrow}c_{\bb k,d\downarrow}\right.\\
    &\qquad\left.+c^\dagger_{\bb k,a\uparrow}c_{\bb k,b\uparrow}\bb q\cdot\nabla_{\bb k}(c^\dagger_{\bb k,c\downarrow}c_{\bb k,d\downarrow})\right)\\
    =&-\sum_{\bb k}A^{abcd}(\bb k)\bb q\cdot\nabla_{\bb k}(c^\dagger_{\bb k,a\uparrow}c_{\bb k,b\uparrow}c^\dagger_{\bb k,c\downarrow}c_{\bb k,d\downarrow})\\
    =&\sum_{\bb k}\bb q\cdot(\nabla_{\bb k}A^{abcd}(\bb k))c^\dagger_{\bb k,a\uparrow}c_{\bb k,b\uparrow}c^\dagger_{\bb k,c\downarrow}c_{\bb k,d\downarrow},
\end{split}
\end{equation}
where at the last step we integrated by part using the fact that the Brillouin zone is compact. Thus there is no contribution to $J(q\rightarrow0)$ as long as $\nabla_{\bb k}A^{abcd}(\bb k)=0$. This is true for the interaction term we used in Eq.~(\ref{HHK}) where $A^{abcd}(\bb k)=U\delta_{ab}\delta_{cd}$. Thus, we can use Eq.~\eqref{eq:current} for the current operator for the purposes of computing response functions in the $q\rightarrow 0$ limit.

\subsection{Hall conductance of QAH-HK model}
 Eq.(\ref{eq:response}) can be directly used to calculate the Hall conductance with the exact Green function Eq.(\ref{eq:green}) using numerical integration techniques. Here we will follow the derivation by Bernevig\cite{bernevig2013topological} by introducing a flat-band limit in order to analytically compute the Hall conductcivity for the QAH-HK model. We place all the occupied energy at $\mu-U<\varepsilon_G<\mu$, whereas all the unoccupied states at energy $\varepsilon_E>\mu$, while keeping the eigenstates of the system unmodified. The Hall conductance (Eq.~(\ref{eq:response})) is invariant under this deformation since the deformation of the dispersion is smooth. We also take the $\beta\rightarrow\infty$ limit to achieve the zero temperature result. 

The non-interacting Hamiltonian for each spin can be written as the sum over the projectors $P_G(\bb k)=\sum_{i\in G}\ket{i,k}\bra{i,k}$ onto the occupied states and the projectors $P_E(\bb k)=\sum_{j\in E}\ket{j,k}\bra{j,k}$ the empty states,
\begin{equation}
    h_\sigma=\varepsilon_GP_G+\varepsilon_EP_E.
\end{equation}
We emphasize that the filled bands are defined according to the interacting system. Thus the projection operators depend on the filling. The current is thus
\begin{equation}
    \frac{\partial h_\sigma(\bb k)}{\partial k_\alpha}=\varepsilon_G\frac{\partial P_G}{\partial k_\alpha}+\varepsilon_E\frac{\partial P_E}{\partial k_\alpha}=(\varepsilon_G-\varepsilon_E)\frac{\partial P_G}{\partial k_\alpha}.
\end{equation}
In the flat band limit, the form of the Green function depends on the filling. We enumerate the possibilities here:
\begin{enumerate}
    \item At quarter-filling, as analyzed above, either spin of the two degenerate $\varepsilon_-$ bands is occupied, thus $\dim{P_G}=2$ and it projects onto the $\varepsilon_-$ bands 
    \begin{equation}
    \begin{split}
        P_Gh&=P_GV\mathrm{diag}(\varepsilon_{G},\varepsilon_{G},\varepsilon_{E},\varepsilon_{E})V^\dagger\\
        &=P_GV\mathrm{diag}(\varepsilon_{G},\varepsilon_{G},0,0)V^\dagger.
    \end{split}
    \end{equation}
    The $\varepsilon_+$ bands remain empty for both spin, $\dim{P_E}=2$ and it projects onto the $\varepsilon_+$ bands.
    \begin{equation}
    \begin{split}
        P_Eh&=P_EV\mathrm{diag}(\varepsilon_{G},\varepsilon_{G},\varepsilon_{E},\varepsilon_{E})V^\dagger\\
        &=P_EV\mathrm{diag}(0,0,\varepsilon_{E},\varepsilon_{E})V^\dagger.
    \end{split}
    \end{equation}
    The exact Green function in the original basis is obtained by performing the unitary transform on Eq.~(\ref{eq:green})
    \begin{equation}
        G(\bb k,\omega)=V\mathrm{diag}(G_{-,\bb k,\uparrow},G_{-,\bb k,\downarrow},G_{+,\bb k,\uparrow},G_{+,\bb k,\downarrow})V^\dagger.
    \end{equation}
    The projection operator $P_G$ ($P_E$) thus projects onto the first two (last two) elements of the diagonal matrix, leaving
    \begin{align}
    \begin{split}
        P_GG(\bb k,\omega)&=V\mathrm{diag}(G_{-,\bb k,\uparrow},G_{-,\bb k,\downarrow},0,0)V^\dagger\\
        &=G_{-,\bb k,\sigma}P_G,
    \end{split}\\
    \begin{split}
        P_EG(\bb k,\omega)&=V\mathrm{diag}(0,0,G_{+,\bb k,\uparrow},G_{+,\bb k,\downarrow})V^\dagger\\
        &=G_{+,\bb k,\sigma}P_E.
    \end{split}
    \end{align}
    Thus the exact Green function at quarter-filling in the orbital %\bbnote{original basis $=$ orbital basis?}
    basis could be written as
    \begin{equation}
        \begin{split}
            G(\bb k,\omega)&=\left(\frac{\frac12}{\omega+\mu-\varepsilon_G}+ \frac{\frac12}{\omega+\mu-U-\varepsilon_G}\right)P_G\\
            &\quad+\frac{1}{\omega+\mu-\varepsilon_E}P_E.
        \end{split}
    \end{equation}
    \item At half-filling, since $U$ is much greater than the bandwidth, the ground state always singly occupies both $\varepsilon_\pm$ bands, this means that both $\varepsilon_+$ and $\varepsilon_-$ are flattened to $\varepsilon_G$. This flattening process does not affect the Hall conductance even if there are zero bands crossing the chemical potential. $P_G$ projects onto all four bands and $P_E=0$ vanishes. The exact Green function in the original basis reads
    \begin{equation}
        \begin{split}
            G(\bb k,\omega)&=\left(\frac{\frac12}{\omega+\mu-\varepsilon_G} + \frac{\frac12}{\omega+\mu-2U-\varepsilon_G}\right)P_G.
        \end{split}
    \end{equation}
    
\end{enumerate}
With these projection operators, the current-current correlator becomes
\begin{equation}
\begin{split}
     &R_{\alpha\beta}(q,i\nu_r)=\frac{1}{V\beta}\sum_{k,n}(\varepsilon_G-\varepsilon_E)^2\\
     &\mathrm{Tr}\left[\frac{\partial P_G}{\partial k_\alpha}G(k+q/2,\omega_n)\frac{\partial P_G}{\partial k_\beta}G(k-q/2,\omega_n-\nu_r)\right].
\end{split}
\end{equation}
One quick inspection of this expression tells us that at half-filling, this response function has to be 0. The derivative on $P_G$ always vanishes as $P_G=I$. Tuning the chemical potential without crossing the poles does not break this 0 value. The Hall conductance at half-filling configuration shall always vanish.

At quarter filling, we adopt the projection operator identities
\begin{align}
    (\partial_\alpha P_G)P_G(\partial_\beta P_G)P_G&=(\partial_\alpha P_G)P_E(\partial_\beta P_G)P_E=0,\\
    (\partial_\alpha P_G)P_G(\partial_\beta P_G)P_E&=(\partial_\alpha P_G)(\partial_\beta P_G)P_E,\\
    (\partial_\alpha P_G)P_E(\partial_\beta P_G)P_G&=(\partial_\alpha P_G)(\partial_\beta P_G)P_G.
\end{align}
Together with the contour integration method to perform the summation over the Matsubara frequencies $\omega_n$, we find that
\begin{equation}
\begin{split}
     R_{\alpha\beta}(i\nu_r)&=\frac{1}{2V}\sum_{k}(\varepsilon_G-\varepsilon_E)^2\\
     &\mathrm{Tr}\left[\frac{(\partial_\alpha P_G)(\partial_\beta P_G)P_E}{\varepsilon_G-\varepsilon_E-i\nu_r}+\frac{(\partial_\alpha P_G)(\partial_\beta P_G)P_G}{\varepsilon_G-\varepsilon_E+i\nu_r}\right].
\end{split}
\end{equation}
The dependence on $U$ is fully removed since the poles at $\varepsilon_G+U$ always lie on the same side as $\varepsilon_E$ relative to the chemical potential. We may complete the integral in the lower half plane without enclosing any poles.
Taking the antisymmetric part between $\alpha$ and $\beta$ and performing the analytic continuation on Matsubara frequencies $i\nu_r\rightarrow\omega$ leads to
\begin{equation}
\begin{split}
     R_{\alpha\beta}(\omega\rightarrow0)&=\frac{1}{2V}\sum_{k,n}(\varepsilon_G-\varepsilon_E)^2\mathrm{Tr}\left[\frac{2\omega(\partial_\alpha P_G)(\partial_\beta P_G)P_G }{(\varepsilon_G-\varepsilon_E)^2-\omega^2}\right]\\
     &=\frac{\omega}{V}\sum_{k,n}\mathrm{Tr}\left[(\partial_\alpha P_G)(\partial_\beta P_G)P_G\right].
\end{split}
\end{equation}
This response function is exactly half of the non-interacting value due to the halved weight in the non-interacting Green function. By substituting the wave function formalism for the projection operator $P_G=\sum_{i\in G}\ket{i,k}\bra{i,k}$, we conclude that the Hall conductance, in units of $\frac{e^2}{h}$, is half of the Berry curvature of the filled non-interacting band, which means $C_1=\frac{1}{2}C_{1}^{\text{non-interacting}}=1$.

\subsection{Ferromagnetic ground state}
The ground state of the HK model is known to possess a large degeneracy due to the spin degrees of freedom\cite{HK,HKnp1}. This degeneracy can be removed by applying an infinitesimal Zeeman field that picks one certain direction for the ferromagnetic ground state\cite{rghk,mai2023topological,mai20221}. By applying an infinitesimal magnetic field along the z-direction, the Green function is modified at all energies. The locations of the poles do not move, but the weight of the Hubbard band poles is now unity, and the bands are fully spin-polarized. At quarter- and half- filling, the system remains insulating, and the exact Green function in the band basis ~\cite{yang2021exactly}
\begin{align}
    G_{\pm,\bb k,\uparrow}(\omega)&=\frac{1}{\omega+\mu-\varepsilon_\pm},\\
    G_{\pm,\bb k,\downarrow}(\omega)&=\frac{1}{\omega+\mu-\braket{n_{\bb k\uparrow}}U-\varepsilon_\pm},
\end{align}
where $n_{\bb k\uparrow}=n_{+,\bb k\uparrow}+n_{-,\bb k\uparrow}$ is the total number of filled electrons in the spin-up bands. The system is now equivalent to two separate QAH systems. In the limit of $U\gg W$, only the spin-up bands will be occupied. Thus the ground state is a spin-polarized QAH state. The Hall conductance is unchanged since the Zeeman field is an infinitesimal perturbation. However, since the Zeeman field removes the zeros, it drastically changes $N_3$ even for values of the chemical potential far from the zero bands. At quarter filling, $N_3$ goes from $2$ to $1$, while at half filling $N_3=0$ for all values of the chemical potential. This shows that $N_3$ is not a property of the ground state manifold but instead depends on the properties of the Green function at all energies. This is in contrast to the Hall conductance which is related to a zero frequency response function.

\end{document}